\documentstyle[12pt]{article}
\newcommand{\newsubsection}[1]{
\vspace{1cm}
\pagebreak[3]
\addtocounter{subsection}{1}
\addcontentsline{toc}{subsection}{\protect
\numberline{\arabic{section}.\arabic{subsection}}{#1}}
\noindent{ \sc  #1}                 
\nopagebreak
\vspace{2mm}
\nopagebreak}
\renewcommand{\theequation}{
\arabic{equation}}

\newlength{\extraspace}
\newlength{\extraspaces}
\newcommand{\be}{\begin{equation}
\addtolength{\abovedisplayskip}{\extraspaces}
\addtolength{\belowdisplayskip}{\extraspaces}
\addtolength{\abovedisplayshortskip}{\extraspace}
\addtolength{\belowdisplayshortskip}{\extraspace}}
\newcommand{\ee}{\end{equation}}
\newcommand{\ba}{\begin{eqnarray}
\addtolength{\abovedisplayskip}{\extraspaces}
\addtolength{\belowdisplayskip}{\extraspaces}
\addtolength{\abovedisplayshortskip}{\extraspace}
\addtolength{\belowdisplayshortskip}{\extraspace}}
\newcommand{\ea}{\end{eqnarray}}
\newcommand{\nonu}{\nonumber \\[2mm]}
\newcommand{\is}{& \!\! = \!\! &}
\newcommand{\half}{{\textstyle{1\over 2}}}

\newcommand{\Z}{{\bf Z}}           
\newcommand{\cH}{{\cal H }}

\newcommand{\ra}{\rightarrow}

\newcommand{\Lbar}{{\overline{L}}}

\def\a{\alpha} 
\def\b{\beta}

\def\adot{{\dot\a}}
\def\bdot{{\dot\b}}
\def\th{\theta}
\newcommand{\R}{{\bf R}}
\renewcommand{\d}{{\partial}}

\newcommand{\Sym}{S}

\newcommand{\ext}{{\raisebox{.2ex}{$\textstyle \bigwedge$}}}

\newcommand{\Dslash}{{D \hspace{-7.4pt} \slash}\;}
\newcommand{\ttr}{\! {\rm tr}\, }
\newcommand{\tr}{{\rm tr}}

\input epsf
\newcounter{fignum}
\newcommand{\figuurnum}{\arabic{fignum}}

\newcommand{\figuurplus}[3]{
\addtocounter{fignum}{1}
\addcontentsline{lof}{figure}{\protect
\numberline{\arabic{section}.\arabic{fignum}}{#3}}
\hspace{-3mm}{\it fig.}\ \figuurnum.
\begin{figure}[t]\begin{center}
\leavevmode\hbox{\epsfxsize=#2 \epsffile{#1.eps}}\\[5mm]
\parbox{10cm}{\small \bf Fig.\ \figuurnum: \it #3}
\end{center} \end{figure}\hspace{-1.5mm}}

\begin{document}

\thispagestyle{empty}

\begin{flushright}
{\sc March 1997}\\
{\sc hep-th/9703030}\\
{\sc cern-th/97-34}\\
{\sc thu-97/06}\\
{\sc utfa-97/06}\\
\end{flushright}

\begin{center}
{\Large\sc{Matrix String Theory}}\\[13mm]

{\sc Robbert Dijkgraaf}\\[2.5mm]
{\it Department of Mathematics}\\
{\it University of Amsterdam, 1018 TV Amsterdam}\\[4mm]

{\sc Erik Verlinde}\\[2.5mm]
{\it TH-Division, CERN, CH-1211 Geneva 23}\\[1mm]
and\\[.1mm]
{\it Institute for Theoretical Physics}\\
{\it Universtity of Utrecht, 3508 TA Utrecht}\\[3mm]

and\\[2mm]

{\sc  Herman Verlinde}\\[2.5mm]
{\it Institute for Theoretical Physics}\\
{\it University of Amsterdam, 1018 XE Amsterdam} \\[1.8cm]

{\sc Abstract}

\end{center}

\noindent
Via compactification on a circle, the matrix model of M-theory
proposed by Banks et al suggests a concrete identification between the
large $N$ limit of two-dimensional ${\cal N}=8$ supersymmetric Yang-Mills
theory and type IIA string theory. In this paper we collect evidence
that supports this identification. We explicitly identify the
perturbative string states and their interactions, and describe the
appearance of D-particle and D-membrane states.

\vfill

\newpage

\newsubsection{Introduction}

By definition, M-theory is the 11-dimensional theory that via
compactification on a circle $S^1$ is equivalent to ten dimensional
type IIA string theory \cite{witten,schwarz}.  The string coupling
constant $g_s$ emerges in this correspondence as the radius of the
$S^1$, while the particles with non-zero KK-momentum along the $S^1$
are identified with the D-particles of the IIA model.  According to
the matrix theory proposal put forward in \cite{banks}, the
full dynamics of M-theory can be captured by means of an appropriate
large $N$ limit of supersymmetric matrix quantum mechanics.

In the original correspondence with the type IIA string, the matrix
degrees of freedom find their origin in the collective dynamics of the
D-particles \cite{polch,bound,ulf}. The key new ingredient in the approach of
\cite{banks}, however, is that the large $N$ limit
effectively accomplishes a decompactification of the extra 11th
direction, which therefore should be treated on the exact same footing
as all the other uncompactified dimensions.  This insight, albeit
still conjectural, provides a number of important new theoretical
tools. Namely, by interchanging the original role of the 11-th
direction with that of one of the other directions, one in principle
achieves a concrete identification of the complete non-perturbative
particle spectrum of string theory in a given dimension with a
relatively convenient subset of states (namely all states that can be
made up from infinitely many D-particles) of string theory
compactified to one dimension less. In this way, much of the recently
developed D-brane technology is upgraded by one dimension and has
become directly applicable in the study of M-theory compactifications.

In this paper we aim to elaborate this new viewpoint for 
the simplest compactification of M-theory, namely on $S^1$.  
Following the current approach in matrix theory \cite{banks,wati}, 
the compactification on this $S^1$ is achieved by reinterpreting the 
infinite dimensional matrices $X^i$ as covariant derivatives $D_i$ 
(written in a Fourier mode basis) of a large $N$ gauge field defined on
the $S^1$.  In the original D-particle language, this procedure 
in fact amounts to applying a T-duality transformation along the 
$S^1$-directions, thereby turning the D-particles into D-strings.
Adopting this approach, we have cast matrix theory into the form
of a two-dimensional ${\cal N}=8$ 
supersymmetric $U(N)$ Yang-Mills theory with the Lagrangian\footnote{Here 
we work in string units $\alpha'=1$. A derivation of (\ref{sym})
from matrix theory and a discussion of our normalizations
is given in the appendix.} 
\be
\label{sym}
S = {1\over 2\pi } \int \tr\left((D_\mu X^i)^2 + \theta^T \Dslash \theta +
g_s^2 F_{\mu\nu}^2 - {1\over g_s^2}[X^i,X^j]^2 + 
{1\over g_s} \theta^T\gamma_i [X^i,\theta]\right).
\ee
Here the 8 scalar fields $X^i$ are $N\times N$ hermitian matrices, as
are the 8 fermionic fields $\theta^\a_L$ and $\theta^{\dot\a}_R$.
The fields $X^i$, $\theta^\a$, $\th^{\dot\a}$ transform respectively 
in the ${\bf 8}_v$ vector, and ${\bf 8}_s$ and ${\bf 8}_c$ spinor 
representations of the $SO(8)$ R-symmetry group of transversal rotations.  
The two-dimensional world-sheet is taken to be a cylinder parametrized 
by coordinates $(\sigma,\tau)$ with $\sigma$ between $0$ and $2\pi$. 
The fermions are taken in the 
Ramond sector, and there is no projection on particular fermion number.
According to the matrix model philosophy, we will consider this
theory in the limit $N \rightarrow \infty$. 

Note that the same set of fields feature in the Green-Schwarz
light-cone formalism of the type II superstring, except that here they
describe non-commuting matrices \cite{banks}. Indeed, the eigenvalues
of the above matrix coordinates $X^i$ are from our point of view
identified with the coordinates of the fundamental type IIA string,
since, relative to the original starting point of large N D-particle
quantum mechanics on $S^1\times R^8$, we have applied an 11-9 flip
that interchanges the role of the 11-th and 9-th directions of
M-theory. The interpretation of $N$ in this matrix string theory is as
(proportional to) the total light-cone momentum $p_+$, see
\figuurplus{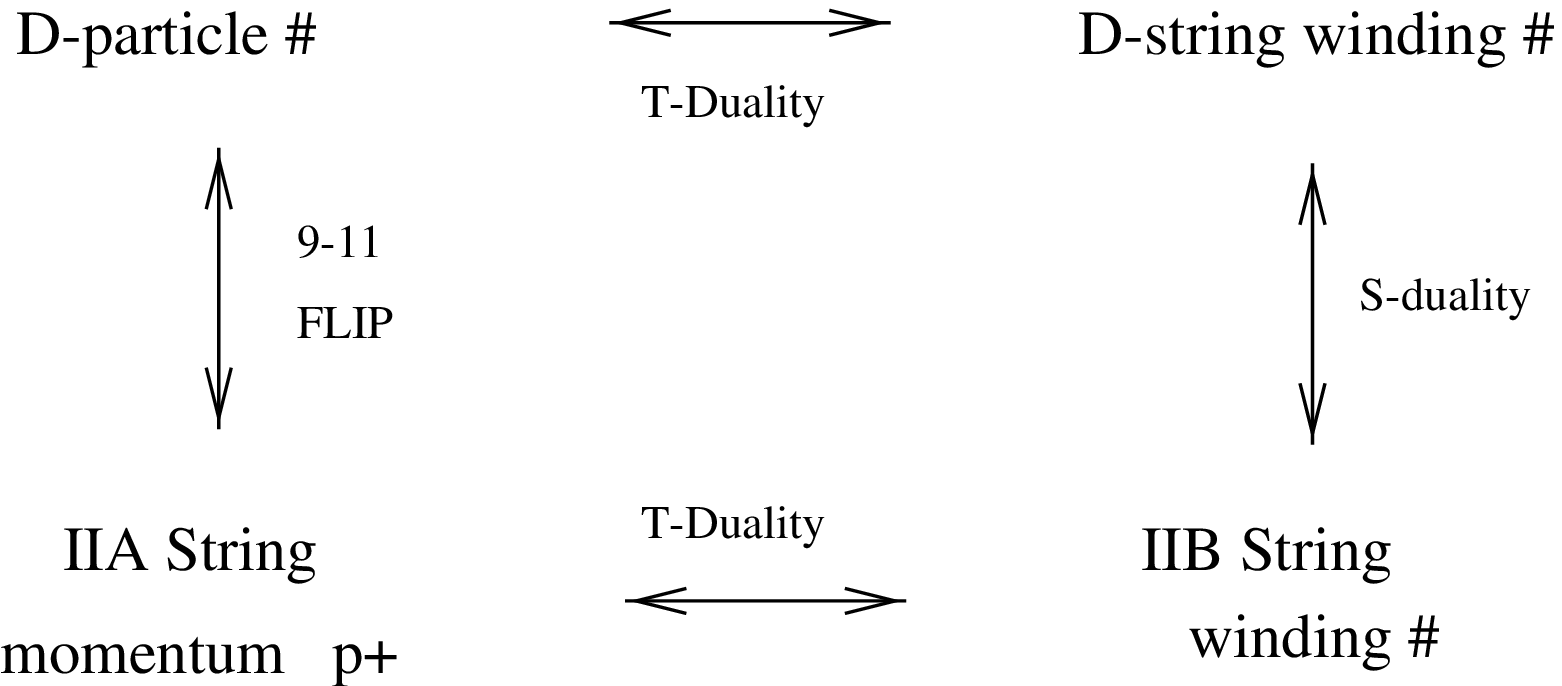}{11cm}{The duality diagram that relates the matrix
string theory to large N D-particle quantum mechanics and the
interpretation of $N$ in the models.}

In the following sections we will elaborate this correspondence in
some detail. We will begin with identifying the complete perturbative
string spectrum, which arises in the IR limit of the two-dimensional
SYM theory.  This correspondence is based on the exact equivalence
between a (free) second-quantized string spectrum and the spectrum
of a two-dimensional $S_N$ orbifold sigma-model\footnote{This new
representation of second quantized string theory was first pointed 
out in \cite{bps} and elaborated in more detail in \cite{orbifold}.}.
We will then proceed to study the string interactions from the
SYM point of view. Finally, we will describe how D-particles and
other D-branes naturally appear in this theory.

While this paper was being written, the preprint \cite{banks-seiberg}
appeared, in which closely related results are reported.  We also
became aware of the earlier work \cite{motl}, where the formulation of
nonperturbative string theory by means of two-dimensional SYM theory
was first proposed and where some of our results were independently
obtained. This approach was also anticipated in \cite{banks-jerusalem,sethi}.
Similar points of view relating strings and matrices are advocated in,
among others, \cite{kawai,wu-etal}.

\newsubsection{The free string limit}

In the interpretation of the ${\cal N}=8$ SYM model as a matrix
string, the usual YM gauge coupling (which has dimension 1/length in
two dimensions) is given in terms of the string coupling as
$g^{-2}_{YM}=\a'g_s^2$. The dependence on the string coupling constant
$g_s$ can be absorbed in the area dependence of the two-dimensional
SYM model. In this way $g_s$ scales inversely with world-sheet length.
The free string at $g_s=0$ is recovered in the IR limit.  In this IR
limit, the two-dimensional gauge theory model is strongly coupled and
we expect a nontrivial conformal field theory to describe the IR fixed
point.

Rather standard argumentation determines that the conformal field
theory that describes this IR limit is the ${\cal N} =8$
supersymmetric sigma model on the orbifold target space
\be
S^N\R^8 = (\R^8)^N/S_N, 
\ee
see e.g. \cite{jeffetal}.
First we observe that in the $g_s=0$ limit the fields $X$ and $\theta$ will
commute. This means that we can write the matrix coordinates as
\be
X^i =  U x^i U^{-1}
\ee
with $U\in U(N)$ and $x^i$ a diagonal matrix with eigenvalues
$x_1^i,\ldots, x_N^i$. That is, $x^i$ takes values in the Cartan 
subalgebra of $U(N)$. This leads to a description of the model in terms
of $N$ Green-Schwarz light-cone coordinates 
$x^i_I,\th^\a_I,\th^\adot_I$ with $I=1,\ldots,N$.

The complete correspondence with a free second-quantized string
Hilbert space in the $N\ra\infty$ limit involves the twisted sectors.
The only gauge invariant quantity is the set of eigenvalues of the
matrices $X^i$.  Therefore, if we go around the space-like $S^1$ of the
world-sheet, the eigenvalues can be interchanged and the fields
$x^i_I(\sigma)$ can be multivalued. Specifically, we have to allow for
configurations with
\be
x^i(\sigma + 2\pi) = g x^i(\sigma) g^{-1},
\ee
where the group element $g$ takes value in the Weyl group of $U(N)$,
the symmetric group $S_N$. The eigenvalue fields $x^i_I(\sigma)$ thus
take values on the orbifold space $S^N\R^8$. These twisted
sectors  are depicted in 
\addtocounter{fignum}{1}
{\it fig.}\ \figuurnum\
and correspond to configurations with strings with various lengths.
\begin{figure}[t]\begin{center}
\leavevmode\hbox{\epsfxsize=6cm \epsffile{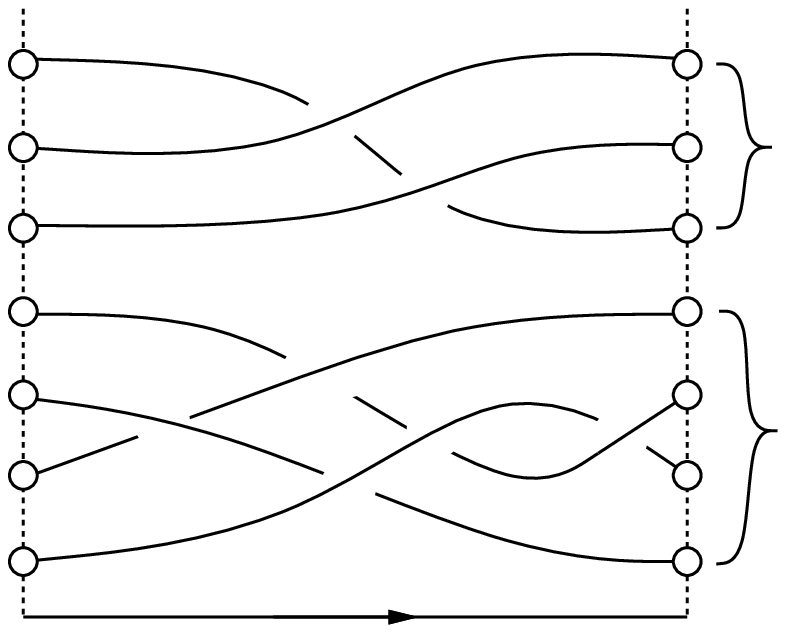}}\\
$0$ \hspace{22mm} $\sigma$ \hspace{22mm} $2\pi$ \\[5mm]
\parbox{10cm}{\small \bf Fig.\ \figuurnum: \it 
A twisted sector 
corresponds to a configuration with strings of various lengths.}
\end{center} \end{figure}\hspace{-1.5mm}

The Hilbert space of this $S_N$ orbifold field theory
is decomposed into twisted sectors labeled by the conjugacy classes
of the orbifold group $S_N$  \cite{orbifold},
\be
\cH(S^N\R^8) = \bigoplus_{{\rm partitions} \ \{N_n\}} \cH_{\{N_n\}}.
\ee
Here we used that for 
the symmetric group, the conjugacy classes $[g]$  are characterized
by partitions $\{N_n\}$ of $N$
\be
\sum_{n} n N_n = N,
\ee
where $N_n$ denotes the multiplicity of the cyclic permutation $(n)$
of $n$ elements in the decomposition of $g$
\be
[g] = (1)^{N_1}(2)^{N_2} \ldots (s)^{N_s}.
\ee
In each twisted sector, one must further keep only the
states invariant under the centralizer subgroup $C_g$ of $g$,
which takes the form
\be
C_g = 
\prod_{n=1}^s \, S_{N_n} \times \Z^{N_n}_n
\ee
where each factor $S_{N_n}$ permutes the $N_n$ cycles $(n)$, while each
$\Z_n$ acts within one particular cycle $(n)$.

Corresponding to this factorisation of $[g]$, we can decompose each
twisted sector into the product over the subfactors $(n)$ of
$N_n$-fold symmetric tensor products of appropriate smaller Hilbert
spaces $\cH_{(n)}$
\be
\label{hdecomposition}
{\cH}_{\{N_n\}} = \bigotimes_{n>0} \, \Sym^{N_n} \cH_{(n)}
\ee
where\footnote{Here 
the symmetrization is assumed to be compatible with the grading
of $\cH$. In particular for pure odd states $S^N$
corresponds to the exterior product $\ext^N$.}
\be
\Sym^N\cH = \left(\underbrace{\cH \otimes \ldots \otimes \cH}_{N\ times}
\right)^{S_N}.
\ee
The spaces $\cH_{(n)}$ in (\ref{hdecomposition}) denote
the $\Z_n$ invariant subsector of the space of states of a
single string on $\R^8\times S^1$ with winding number $n$. We can
represent this space via a sigma model of $n$
coordinate fields $x_I^i(\sigma) \in X$ with the cyclic boundary
condition
\be
\label{boundaryc}
\qquad x_I^i(\sigma+2\pi) = x_{I+1}^i(\sigma), \qquad I \in (1, \ldots, n).
\ee
We can glue the $n$ coordinate fields $x_I(\sigma)$ together into one
single field $x(\sigma)$ defined on the interval $0\leq
\sigma \leq 2\pi n$. Hence, relative to the string with winding number
one, the oscillators of the long string that generate $\cH_{(n)}$ have a
fractional  ${1\over n}$ moding. 
The group $\Z_n$ is generated by the cyclic permutation
\be
\label{omega}
\omega : \ x_I \rightarrow x_{I+1}
\ee
which via (\ref{boundaryc}) corresponds to a translation $\sigma \ra 
\sigma + 2\pi$. Thus the $\Z_n$-invariant subspace consists of those states 
for which the fractional left-moving minus right-moving oscillator numbers 
combined add up to an integer. 

It is instructive to describe the implications of this structure
for the Virasoro generators $L_0^{(i)}$ of the individual strings.
The total $L_0$ operator of the $S_N$ orbifold CFT in a twisted
sector given by cyclic permutations of length $n_i$ decomposes
as
\be
L^{tot}_0 = \sum_i {L_0^{(i)}\over n_i}.
\label{decomp}
\ee
Here $L_0^{(i)}$ is the usual canonically normalized operator in terms
of the single string coordinates $x(\sigma),\th(\sigma)$ defined above. 
The meaning of the above described $\Z_n$ projections is that it requires 
that the contribution from a single string sector to the total world-sheet
translation generator $L{}^{tot}_0-\Lbar{}^{tot}_0$ is integer-valued. 
From the individual string perspective, this means that 
$L{}^{(i)}_0-\Lbar{}^{(i)}_0$ is a multiple of $n_i$.

To recover the Fock space of the second-quantized type IIA string we
now consider the following large $N$ limit. We send $N\ra \infty$ and
consider twisted sectors that typically consist of a finite number of
cycles, with individual lengths $n_i$ that also tend to infinity in
proportion with $N$. The finite ratio $n_i/N$ then represents the
fraction of the total $p_+$ momentum (which we will normalize to 
$p_+^{tot}=1$) carried by the corresponding string\footnote{The fact that
the length $n_i$ of the individual strings specifies its light-cone
momentum is familiar from the usual light-cone formulation of 
string theory \cite{GSW}.}
\be
p_+^{(i)}= {n_i\over N}.
\label{p+}
\ee
So, in the above terminology, only long strings survive. The
usual oscillation states of these strings are generated in the orbifold CFT 
by creation modes $\a_{-k/N}$ with $k$ finite. Therefore, in the large 
$N$ limit only the very low-energy IR excitations of the $S_N$-orbifold
CFT correspond to string states at finite mass levels. 

The $\Z_n$ projection discussed above in this limit effectively
amounts to the usual uncompactified level-matching conditions $L{}^{(i)}_0 -
\Lbar{}^{(i)}_0=0$ for the {\it individual} strings, since all single string
states for which $L_0-\overline{L}_0 \not=0$ become infinitely massive at
large $N$. The total mass-shell condition reads (here we put the string
tension equal to $\alpha' = 1$)
\be
p_-^{tot}=NL_0^{tot}
\ee
and we recover the mass-shell conditions of the individual strings by 
decomposing $L_0^{tot}$ as in (\ref{decomp}), via the definition of the 
individual $p_-$ light-cone momentum as
\be
p_-^{(i)}= {NL_0^{(i)}\over n_i}
\ee
which combined with (\ref{p+}) 
gives the usual relation $p_+^{(i)}p_-^{(i)}=L_0^{(i)}$. All strings
therefore indeed have the same string tension.

\newsubsection{Interactions}

We have seen that the super-Yang-Mills model in the IR limit gives the
free light-cone quantization of the type IIA string in terms of an
orbifold sigma model, that describes the completely broken phase
of the $U(N)$ gauge symmetry. The twisted sectors of the orbifold describe
multi-string states and are superselection sectors in the
non-interacting model --- the number of strings in conserved for
$g_s=0$.

The conservation of string number and of individual string momenta
will be violated if we turn on the interactions of the 1+1-dimensional
SYM theory. The idea is that by relaxing the strict IR limit, one
gradually needs to include configurations in which the non-abelian
symmetry gets restored in some small space-time region.  Indeed, if at
some point in the $(\sigma,\tau)$ plane two eigenvalues $x_I$ and
$x_J$ coincide, we enter a phase where an unbroken $U(2)$ symmetry is
restored, and we should thus expect that for non-zero $g_s$, there
will be a non-zero transition amplitude between states that are
related by a simple transposition of these two
eigenvalues\footnote{This observation was first made in
\cite{motl}.}. In the IR $S_N$ orbifold theory, such a process will
correspond to a local interaction on the two-dimensional world-sheet,
which can be seen to correspond to the elementary joining and
splitting of strings.

To see this more explicitly, consider a
configuration that connects two different sectors, labeled by $S_N$
group elements that are related by multiplication by a simple
transposition. It is easy to see that this simple 
transposition connects a state with, say, one string represented by a 
cycle $(n)$ decays into a state with two strings represented by a 
permutation that is a product of two cycles $(n_1)(n_2)$ with 
$n_1+n_2=n$, or vice versa. So the numbers of incoming and outgoing 
strings differ by one. Pictorially, what takes place is that
the two coinciding eigenvalues connect or disconnect at the intersection 
point, and as illustrated in \figuurplus{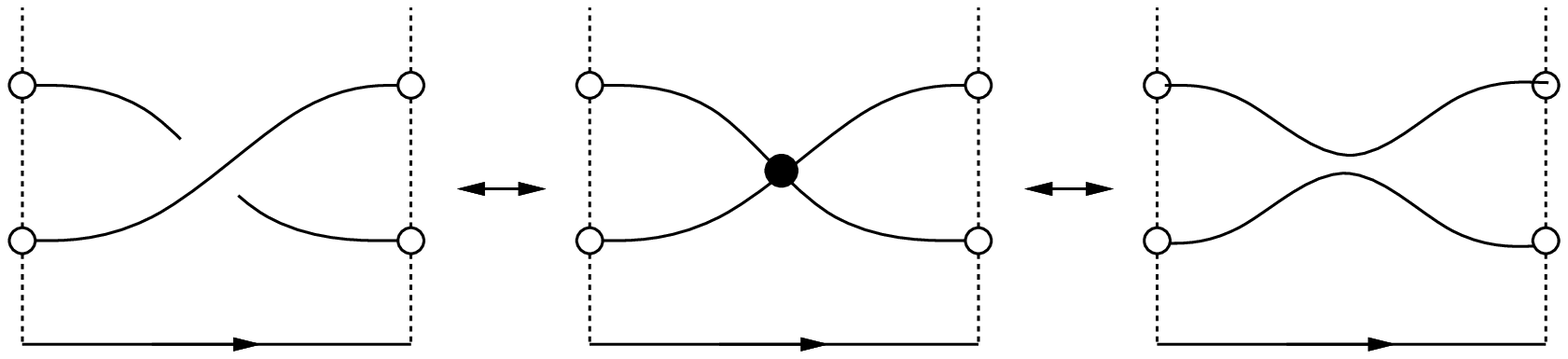}{12cm}{The splitting 
and joining of strings occurs if two eigenvalues coincide.}
this indeed represents an elementary string interaction.

In the CFT this interaction is represented by a local operator, and thus
according to this physical picture one may view the SYM theory as obtained 
via a perturbation of the $S_N$-orbifold conformal field theory. In first 
order, this perturbation is described via a modification of the CFT action 
\be
\label{llambda}
S = S_{CFT} + \lambda \int\! d^2z V_{int}
\ee
where $V_{int}$ is an appropriate (e.g. space-time supersymmetric)
twist operator, that generates the just described simple transposition
of eigenvalues. We now claim that this joining and splitting process
is indeed first order in the coupling constant $g_s$ as defined in the
SYM Lagrangian (\ref{sym}). Instead of deriving this directly in the
strongly coupled SYM theory, we analyze the effective operator that
produces such an interaction in the IR conformal field theory.
The identification of $\lambda$ in (\ref{llambda}) in terms of
the coupling constants in (\ref{sym}) via
\be
\label{scaling}
\lambda \sim g_s \sqrt{\a'}
\ee 
requires that this local interaction vertex $V_{int}$ must have
scale dimension $({3\over 2},{3\over 2})$ under scale transformations
on the two-dimensional world-sheet. We will now verify that this is
indeed the case.

\newsubsection{The interaction vertex}

It is clear from the above discussion that the interaction vertex
$V_{int}$ will be a twist field that interchanges two eigenvalues, say
$x_1$ and $x_2$. It acts therefore as a $\Z_2$ reflection on the
relative coordinate $x_1-x_2$. This coordinate has 8 components, so
the total conformal dimension of the corresponding superconformal
field (including the fermionic contribution) is $(1,1)$. The
corresponding descendent operator that one can include in the action
has dimensions $({3\over 2},{3\over 2})$.  Note that this is an
irrelevant operator, that disappears in the IR limit. In fact, the
corresponding coupling constant has dimension $-1$. Since the
world-sheet length-scale is inversely proportional to the string
coupling $g_s$, we immediately see that this interaction linear in
$g_s$, which is what we set out to establish. In fact, we will show
that the twist field is uniquely characterized as the least irrelevant
operator in the CFT that is both space-time supersymmetric and Lorentz
invariant.

Let us do this computation in a bit more detail. First we recall that
in the CFT description of the second-quantized GS light-cone string
the spacetime supercharges are given by (in the normalization
$p_+^{tot} = 1$)
\be
Q^\a = {1\over \sqrt{N}}\oint \!d\sigma\,  \sum_{I=1}^N \th_I^\a,\qquad
Q^\adot =\sqrt{N} \oint \!d\sigma\, G^\adot,
\ee
with
\be
\qquad G^\adot(z) = \sum_{I=1}^N \gamma^i_{\a\adot} \th_I^a \d x_I^i
\ee
the generators of the total ${\cal N}=8$ world-sheet supersymmetry
algebra, with similarly defined right-moving charges.

Concentrating on the left-moving sector, let us analyze the twist
fields of the $\Z_2$ twist that interchanges two eigenvalues, say
$x_1$ with $x_2$ and similarly $\th_1$ and $\th_2$. One first goes
over to the eigenvectors $x_\pm=x_1\pm x_2$, $\th_\pm=\th_1\pm\th_2$.
The $\Z_2$ acts on the minus components as
\be
x_-^i,\th_-^\a \ra -x_-^i,-\th_-^\a,
\ee
so we are essentially dealing with a standard $\R^8/\Z_2$ supersymmetric
orbifold.

This orbifold has well-known twist fields, which
will receive contributions of the bosons and fermions.
The bosonic twist operator $\sigma$ is defined via the
operator product
\be
x_-^i(z) \cdot \sigma(0) \sim z^{-{1\over 2}} \tau^i(0)
\label{ope1} 
\ee
Since the fermions transform in spinor representation ${\bf 8}_s$, their
spin fields $\Sigma^i,\Sigma^\adot$ will transform in the vector
representation ${\bf 8}_v$ and the conjugated spinor representation
${\bf 8}_c$. They are related via the operator products
\ba
\th_-^\a(z) \cdot \Sigma^i(0) & \sim & 
z^{-{1\over 2}} \gamma^i_{\a\adot}\Sigma^\adot(0)  \nonu
\th_-^\a(z) \cdot \Sigma^\adot(0) & 
\sim & z^{-{1\over 2}} \gamma^i_{\a\adot}\Sigma^i(0)
\ea
The bosonic twist field $\sigma$ and the spin fields 
$\Sigma^i$ or $\Sigma^\adot$ have all conformal dimension $h=\half$ (i.e.
${1\over 16}$ for each coordinate), and the conjugated field $\tau^i$ 
in (\ref{ope1}) has dimension $h=1$. 

For the interaction vertex we propose the following
space-time supersymmetric, $SO(8)$ invariant, weight ${3\over 2}$ field
\be
\tau^i\Sigma^i
\ee
This twist field lies in the NS sector of the CFT, and 
represents an interaction between incoming and outgoing Ramond 
states.\footnote{Even though the fermion has Ramond spin structure,
the supercurrent $G^\adot$ obeys NS periodicity.
Using $N=2$ language the chiral primary fields are $\sigma\Sigma^i$ 
and $\sigma\Sigma^\adot$. After spectral flow
they give Ramond ground states.}
It can be written as 
the descendent of the chiral primary field $\sigma\Sigma^\adot$,
since (no summation over $\adot$)
\be
[G^\adot_{-{1\over 2}},\sigma\Sigma^\adot] = \tau^i\Sigma^i.
\ee
This is a special case of the more general identity
\be
[G^\adot_{-{1\over 2}},\sigma\Sigma^\bdot]+ 
[G^\bdot_{-{1\over 2}},\sigma\Sigma^\adot] =
\delta^{\adot\bdot} \tau^i \Sigma^i.
\ee
The interaction vertex operator satisfies
\be 
[G^\adot_{-{1\over 2}},\tau^i\Sigma^i]= \d_z(\sigma\Sigma^\adot),
\ee
as is clear by using the Jacobi identity and the above relation. For the 
full description we also have to include the right-moving degrees of freedom.

To obtain the complete form of the effective world-sheet interaction
term we have to tensor the left-moving and right-moving twist fields
and to sum over the pairs of $I,J$ labeling the two possible
eigenvalues that can be permuted by the $\Z_2$ twist
\be
 \sum_{I<J} 
\lambda \int d^2\!z\; \left(\tau^i\Sigma^i \otimes 
\overline{\tau}^j\overline{\Sigma}^j\right)_{IJ}.
\label{vertex}
\ee
This is a weigth $({3\over 2},{3\over 2})$ conformal field. The
corresponding coupling constant $\lambda$ has therefore total
dimension $-1$ and the interaction will scale linear in $g_s$
just as needed, see eqn (\ref{scaling}).  

The interaction is space-time supersymmetric. First, it
preserves the world-sheet ${\cal N}=8$ supersymmetry representing the
unbroken charges $Q^\adot$, since supersymmetric variations become
total derivatives and we integrate over the string world-sheet. The
broken space-time supersymmetries $Q^\a$ are trivially conserved.
Although the fermion fields $\th^\a_-$ satisfy Ramond boundary
conditions, and therefore pick up a minus sign in the local coordinate
$z$ when transported around $z=0$, $Q^\a$ is proportional to the
zero-mode of the linear sum $\th_+^\a$, which is not broken by the
twist field interaction. 

It is interesting to compare the above twist field interaction with
the conventional formalism of light-cone string theory\footnote{We
thanks N.\ Berkovits for pointing out this relation to us.}.  As is
discussed by Mandelstam \cite{mandelstam}, in order to obtain a
fully $SO(9,1)$ Lorentz invariant interaction in the light-cone formalism,
it does not suffice to  consider only the geometric joining and
splitting interaction that in our formalism is represented by the
insertion of the twist field $\sigma(0)$.  The interaction has to be
supplemented with a further operator insertion that reads in our
notation
\be
\oint \! {dz \over z^{1\over 2}} \,\Sigma^i \d x^i (z)\, \sigma(0) = 
\tau^i\Sigma^i(0).
\ee
In comparison with the Riemann surfaces picture of interacting strings
\cite{mandelstam}
one should note that around the interaction point the usual string
world-sheet is actually a double cover of the local coordinate $z$ of
our orbifold CFT. The above operator insertion should be tensored with
a similar expression for the right-movers, giving precisely our
interaction vertex (\ref{vertex}). We further remark that the current
$\Sigma^i\d x^i$ has a simple interpretation as the ${\cal N}=1$
world-sheet supercurrent in the covariant NSR formulation of the type
II string. Indeed, in the NSR language the above operator insertion
corresponds simply to picture changing. The fact that the twist field
interactions (and the higher $n$-point vertices obtained through
contact terms) respect the ten-dimensional Lorentz invariance is of
course highly suggestive that the matrix string will also be Lorentz
invariant.

\newsubsection{Hamiltonian formulation}

It will be convenient in the following to represent
the degrees of freedom in a Hamiltonian form, by introducing 
the standard conjugate variables $(\Pi_i,X^i)$ for the scalar fields, 
and $(E,A_1)$, with $E$ the 1-dimensional electric field, for the
gauge fields. The Hamiltonian $H$ and total 1-momentum $P$ 
then take the form
\be
H =\oint \!d\sigma\,  T_{00} \qquad \qquad P= \oint  \!d\sigma\, T_{01}
\ee
where $T_{00}$ and $T_{01}$ are the components
of the two-dimensional energy-momentum tensor, given by
\ba
T_{00} \is
{1\over 2} \tr\Bigl(\Pi_i^2 + (DX_i)^2 + \theta^T \gamma^9 D\theta \nonu
& & \qquad + {1\over g_s} \theta^T\gamma^i[X_i,\theta] + {1
\over g_s^2} \,(E^2 +  [X^i,X^j]^2)\Bigr) \nonumber \\[3mm]
T_{01} \is  \tr\Bigl(\Pi_iDX^i + \theta^T D \theta\Bigr)
\label{momentum}
\ea
Here $D$ is the YM covariant derivative along the spatial direction of the
world-sheet and the $\gamma$-matrices are taken to be $16 \times 16$ matrices.
The spacetime supercharges of the matrix string model take the form
(now $\a$ denotes a non-chiral spinor and runs from 1 to 16)
\be
\label{susys}
\tilde{Q}^\a = {1\over \sqrt{N}} \oint  \!d\sigma\, \tr \th^\a,\qquad
Q^\a = \sqrt{N}\oint  \!d\sigma\, G^\a,
\ee
with
\be
G^\a = \tr\Bigl(\theta^T (\gamma^9E 
+\gamma^{9i} DX_i + \gamma^i\Pi_i + \gamma^{ij}[X_i,X_j])\Bigr)^\alpha 
\ee

\newsubsection{Compactification}

Following the current approach in matrix theory, the compactification of 
the matrix string on a torus $T^d$ is achieved by reinterpreting the 
infinite dimensional matrices $X^i$ as covariant derivatives $D_i$ 
(written in a Fourier mode basis) of a large $N$ gauge field defined 
in these extra dimensions \cite{banks,wati}. 
Adopting this procedure, the matrix string 
becomes equivalent to $d+2$-dimensional supersymmetric gauge field 
theory on the space-like manifold  $S^1\times T^d$. In this correspondence, 
the string coupling constant $g_s$ is identified with square root of the 
volume of the torus $T^d$ (in string units), and thus the free string
limit amounts to shrinking the volume of the extra dimensions inside the 
$T^d$ to zero. In general, like for the uncompactified theory, the 
large $N$ limit of the gauge model needs to be accompanied by an 
appropriate IR-limit in the $S^1$ direction to find correspondence
with the string theoretic degrees of freedom.

For the case $d=2$, in which case the large $N$ supersymmetric gauge
theory lives in 3+1-dimensions, this exact rescaling was in fact
considered previously in \cite{highenergy}, see also
\cite{jeffetal}.\footnote{The weak coupling limit $g_s\rightarrow 0$
was interpreted in \cite{highenergy} as a high energy limit, in which
the typical length scale of excitations in two of the four directions
is taken to be much smaller than in the other two transversal
directions.  It seems worthwhile to further elaborate this
correspondence.}  As also pointed out in
\cite{susskind,ganor}, the implied equivalence of large
$N$ four-dimensional SYM theory to type II string theory compactified
on $T^2$ provides a natural explanation of the $SL(2,Z)$ $S$-duality
symmetry of the former in terms of the $T$-duality of the latter. In
the $g_s \rightarrow 0$ rescaling limit of the gauge theory, this
connection between $S$-duality and $T$-duality of the resulting
$S_N$-orbifold CFT was first pointed out in \cite{jeffetal}.

In general, the maximal number of dimensions one can compactify in
this fashion is eight. In this case matrix theory becomes equivalent
to $9+1$ dimensional large $N$ SYM theory, whose IR behaviour should
describe the type II string compactified to 1+1 dimensions. In
principle one should be able to compactify one more dimension by
taking the light-cone to be a cylinder. The finite $N$ theory may seem
to be a candidate for describing the sector of this compactification
with given discrete momentum $N$ along the extra $S^1$. Although this
procedure is adequate for computing the BPS spectrum of the theory, we
suspects it gives an incomplete description of the dynamics of general
non-BPS configurations.

\newsubsection{Fluxes and Charges}

The above compactification procedure opens up the possibility of adding
new charged objects to the theory, essentially by considering gauge
theory configurations that carry non-trivial fluxes or other topological
quantum numbers along the compactified directions. In particular, 
one can consider configurations with non-zero magnetic fluxes through 
the various two-cycles of the compactification torus $T^d$, and in this
way one can introduce the D-membranes. More generally,
using the correspondence with the matrix theory proposal of 
\cite{banks,branes}, a (partially complete) list of fluxes and 
their type IIA interpretations are listed below
\ba
\mbox{\underline{2D SYM Theory}}\quad &\qquad 
\qquad \qquad  & \qquad \mbox{\underline{type IIA String}} 
\nonumber\\[5mm]
\oint \ttr E \ \qquad  \quad \  & =  &  \qquad \mbox{D-particle $\#$ \ $q_0$}\nonumber\\[2mm]
\oint \ttr \Pi_i \ \qquad  \quad & = & \qquad \mbox{momentum \ $p_i$} \nonumber\\[2mm]
\oint \ttr DX^i \quad  \ \quad & =  & \qquad \mbox{NS winding  \ $w_i$}
\nonumber\\[2mm]
\oint T_{01} \qquad  \quad & = &  \qquad \mbox{NS winding \ $w_+$}
\nonumber\\[2mm]
\oint \ttr [X^i,X^j] \quad \quad & = &  \quad \mbox{D-membrane  \ $\#$ $m_{ij}$}
\nonumber\\[2mm]
\oint T_{0i} \qquad  \quad & = & \quad \mbox{D-membrane  $\#$ \ $m_{i+}$}
\nonumber\\[2mm]
\oint \ttr DX^{[i}X^j X^{k]} \ \  & =  &  \quad \mbox{D-fourbrane $\#$ 
\ $R_{ijk+}$}\nonumber\\[2mm]
\oint \ttr X^{[i}X^jX^kX^{l]} \ \  & =  &  \quad \mbox{NS fivebrane $\#$ 
\ $W_{ijkl+}$}\nonumber\\
\nonumber
\ea
Here $T_{01}$ was defined in equation (\ref{momentum}) and 
$$
T_{0i} = \tr\Bigl( EDX_i + \Pi^j[X_i,X_j] + \theta^T [X_i,\theta]\Bigr)
$$
which in the $d+2$ dimensional SYM language is just the momentum flux 
in the $i$-th direction.

The above topological charges all appear as central terms in the 
supersymmetry algebra generated by the supercharges defined in (\ref{susys}).
One finds \cite{branes}
\ba
\{\tilde Q^\a, \tilde Q^\b\} \is \delta^{\a\b} \nonu
\{Q^\a,\tilde Q^\b \} 
\is (\gamma^9 q_0 + \gamma^ip_i + \gamma^{9i} w_i + \gamma^{ij} m_{ij})^{\a\b}
\\[2mm]   
\{Q^\a,Q^\b\} \is N (H + \gamma^9 w_+  + \gamma^i m_{i+}+ 
\gamma^{9ijk}R_{ijk+} + \gamma^{ijkl} W_{ijkl+})^{\a\b}\nonumber
\ea
Here we recognize the central terms corresponding to the various
charged (extended) objects that are present in the theory.
In the last line, the complete right-hand side is proportional to $N$,
and thus these terms diverge for $N \rightarrow \infty$ as soon as one 
of the central charges $m_{i+}$, $R_{ijk+}$ or $W_{ijkl+}$ is non-zero. 
This corresponds to the fact that the 9th direction is necessarily 
decompactified in the large $N$ limit, and thus these configurations 
represent string-like solitons with an infinite extent in the 9th direction.

\newsubsection{D-particles}

It is not immediately evident that the string, D-brane and fivebrane
configurations as defined above will indeed behave exactly as in
perturbative string theory.  Here we will establish this
correspondence for the case of the D-particle and the D-membrane. In
particular, we will see that they indeed give rise to a new
perturbative sector of strings, that satisfy Dirichlet boundary
conditions on a corresponding codimension subspace.

First we consider a configuration with  D-particle charge equal to $q_0$. 
In the SYM language, this corresponds to a non-zero electric flux.
\be
{1\over 2\pi} \oint  \!d\sigma\, \tr E = q_0.
\ee
This correspondence can be understood in (at least) two ways. First, the
electric flux arises upon compactification and T-duality as the KK-momentum
in the extra 9th dimension. Since this direction was used to compactify
M-theory to ten dimensions, this momentum gets interpreted as D-particle
charge \cite{witten}. Alternatively, as indicated in {\it fig.} 1,
after a T-duality and an S-duality, we can map our IIA strings to the
D-strings of the type IIB theory. As shown in \cite{bound}, fundamental
type IIB strings attach to these D-strings by creating a non-zero electric
flux on the SYM theory that described the D-string world-sheet dynamics.
Inverting the above duality transformations, these fundamental IIB strings
get identified with the D-particles in the IIA setup.

The simplest classical 
configuration that carries such an electric flux $q_0$ is
\be
E = {\bf 1}_{q_0\times q_0}.
\label{flux}
\ee
The presence of the electric flux will break the gauge group as
\be
U(N) \ra U(N\! -\! q_0) \times U(q_0).
\ee
The $U(N-q_0)$ sector, that does not carry an electric flux, represents
the type IIA strings in the background of the D-particles. In the large $N$,
$g_s\ra 0$ limit this gives the usual free closed string spectrum. 
The $U(q_0)$ sector will describe the D-particle degrees of freedom and we
will now examine it in more detail.

We have a similar breaking of the permutation group symmetry from
$S_N$ to $S_{N-q_0} \times S_{q_0}$. The symmetric group $S_{q_0}$
describes the statistics of the D-particles.  The Hilbert subspace
that carries the electric flux $q_0$ will decompose (at least for weak
string coupling) in twisted sectors labeled by partitions of
$q_0$. These sectors have an interpretation as all the possible bound
state configurations of the $q_0$ D-particles.  

The eigenvalues of the $U(q_0)$ part of the matrices $X^i$ can depend 
on the world-sheet parameter $\sigma$ and thus a priori seem to describe 
strings. However, we would like to interprete them as D-particles.
An important point is that in the large $N$ limit we have to keep the
total D-particle charge $q_0$ finite.  This implies that the strings
in the $U(q_0)$ sector become short strings with infinitely massive
oscillations. These short string
oscillations will therefore decouple at large $N$, leaving only their
constant modes. These constant modes describe the positions of the 
D-particles and their various bound states. 
This behaviour should be contrasted with
the eigenvalues in the remaining $U(N\! -\! q_0)$ sector, which can form the
type IIA strings with the usual oscillation spectrum.
Typical configurations thus consist of short
strings describing the D-particles and long type IIA closed strings,
as depicted in \figuurplus{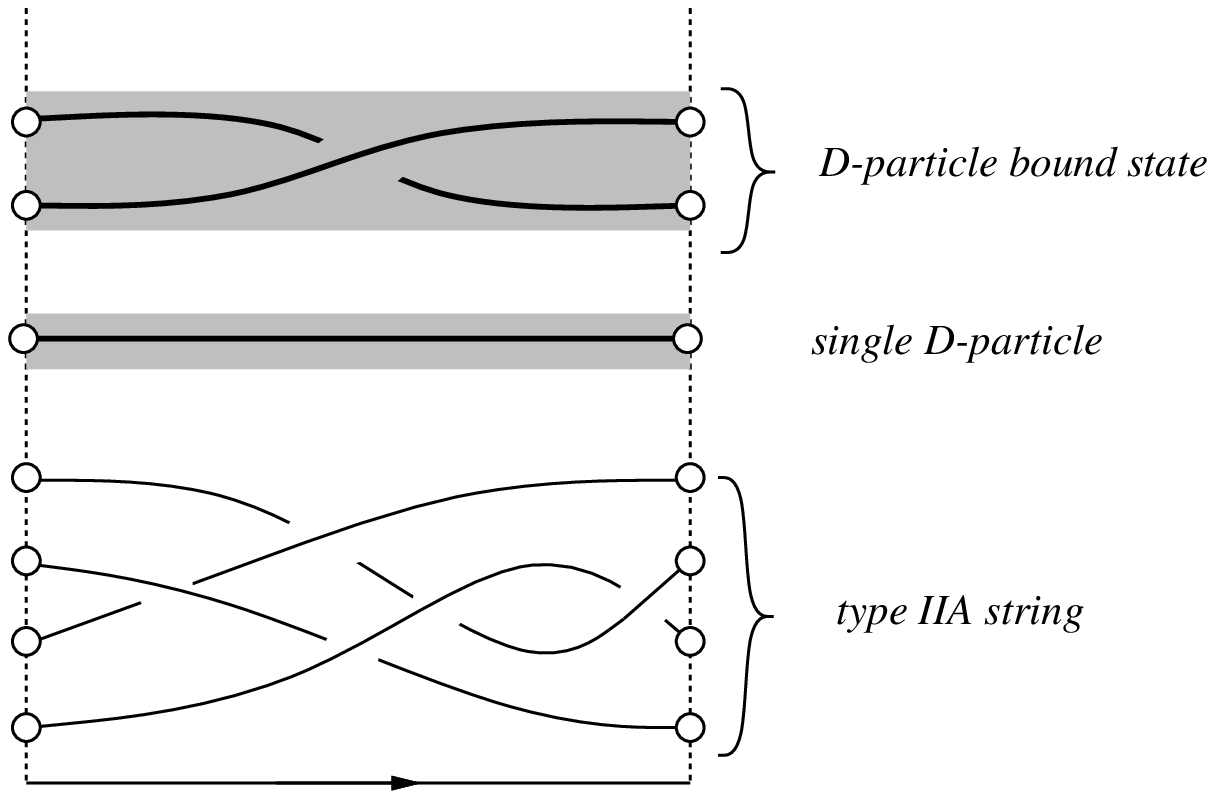}{10cm}{If
we put a flux in a finite rank subgroup of the SYM gauge group, the
corresponding eigenvalues give ``short strings'' that represent the
D-particles and their bound states. The ``long strings'' in the
remaining sector describe the closed type IIA strings in the
background of the D-particles.}

It is worthwhile to note that here we naturally arrive at the existence 
of D-particle bound states, since these automatically arise as twisted 
sectors in the orbifold conformal field theory. So unlike for the original
matrix model based on large $N$ supersymmetric quantum mechanics,
the existence of bound states is not an assumption but a direct
consequence of well-established facts!

An electric flux of the form (\ref{flux}) will give a contribution
$H=q_0/g_s^2$ to the SYM Hamiltonian. In the untwisted sector, that
describes $q_0$ free D-particles, the usual mass-shell relation
gives D-particle masses $m=1/g_s$, which is the expected
result. Similarly, in a twisted sector that describes bound states of
$n_i$ D-particles, with $\sum n_i=q_0$, we find that the bound states
have masses $m_i = n_i/g_s$. This is implied by the mass-shell
relation, quite similarly as in our discussion of the free
strings, since the contribution to the total $L_0$ is of the form
\be
\label{dmass}
L_0^{tot} = {q_0\over g_s^2} + \sum_i {p_i^2\over n_i}
\ee
In particular, for a maximally twisted sector with one cycle of length $q_0$
we find $H=q_0/g_s^2 + p^2/q_0$ which implies $m=q_0/g_s$.

It should be emphasized that, as seen from eqn (\ref{dmass}),
the D-particles states just described in fact have very small
momentum in the light-cone direction. Formally, they have 
$p_+ = {n_i\over N} \rightarrow 0$ in the large $N$ limit. 
It is clear, however, that one can give non-zero longitudinal 
momentum to  the D-particles by attaching them ({\it i.e.} the short 
strings) to a long string.

Do these particles indeed interact with the type II strings
as D-objects? In particular, we would like to see that the theory 
contains a sector of long strings,
which attached to these particles in the usual fashion. 
The natural candidate for such configurations are twisted sectors in
which (in spite of the symmetry breaking due to the presence of the
electric flux) two different kind of eigenvalues are transposed,
corresponding to respectively the long type IIA strings and
the short string that represent the D-particles. In other words,
the topology of the eigenvalues of the $X^i$ fields in this sector is
incompatible with topology of the eigenvalues of the (non-abelian)
electric field $E$, in that the direction of the $E$ field in the Lie
algebra must necessarily point outside of the Cartan subalgebra
specified by the $X^i$ fields (and vice versa).  In the IR limit this
would however be energetically unfavorable, {\it unless} the short and 
long eigenvalue (that are transposed along the $S^1$) coincide. Where this
happens the unbroken gauge group gets locally enhanced to $U(2)$, and
this allows the eigenvalues to cross without much loss of energy, see
\figuurplus{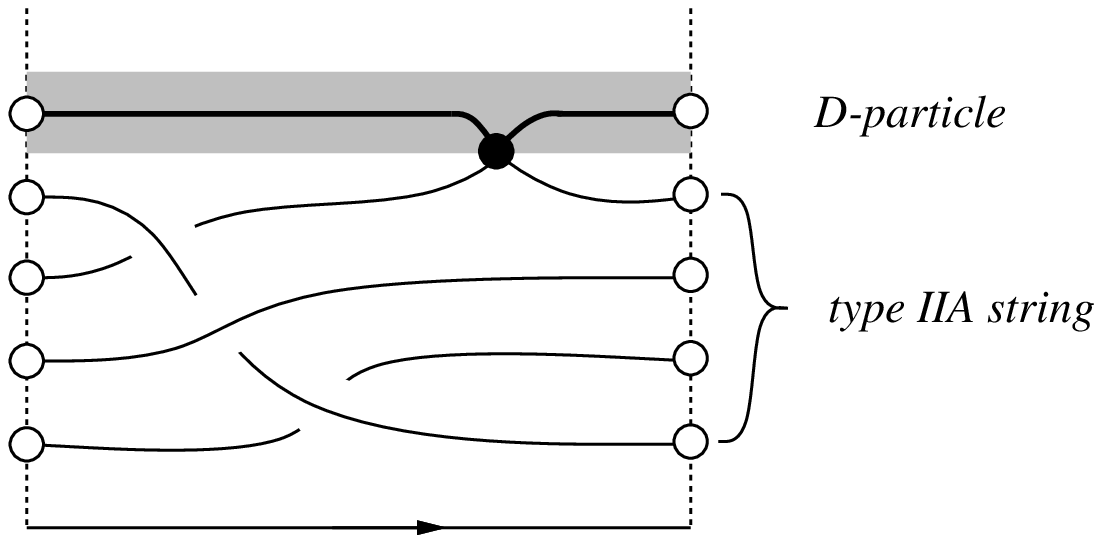}{9cm}{In perturbation theory the closed string can
get glued to the D-particle.}  In this way, the long string eigenvalue
indeed stays glued to the short string eigenvalue representing the
D-particle.

The ground state of such a bound state between a long and a short string
represents the D-particle moving with non-zero $p_+$. A quantitative 
verification of this picture is as follows. A long string of total
length $n$ occupies a $U(n)$ subgroup. If the string is in its ground 
state, all $x_I$ essentially coincide and thus the $U(n)$ gauge symmetry
is unbroken. If the configuration further carries an electric flux $q_0$, 
this flux can lower its worldsheet energy by spreading into this
unbroken gauge group\footnote{We thank T. Banks for suggesting this picture.}.
The flux then takes the form of a flux ${q_0 \over n}$ inside the diagonal 
$U(1)$, together with an opposite 't Hooft type electric flux 
${q_0\over n}$ inside the $SU(n)$. The energy of this flux 
is now smaller by a factor of $n$, in exact accordance with the fact 
it now corresponds to a D-particle with finite $p_+ = n/N$.

These arguments are admittedly somewhat qualitative. It would be useful
to perform a more quantitative study that supports the presented
physical picture.

\newsubsection{D-membranes}

D-membranes are configurations with non-zero values for the topological
charge
\be
m_{ij} = \tr[X_i,X_j].
\ee
Let us briefly recall the construction of such configurations in 
the matrix model of M-theory 
and the resulting correspondence with the membrane 
world-volume theory. In $U(N)$ one can find two matrices $U$ and $V$, 
such that 
$$
UV =e^{2\pi i\over N} VU.
$$
Any hermitian matrix $X$ can then be written as
$$
X = \sum_{nm} x_{nm} U^n V^m
$$
and this expansion can be used to associate to $X$ the function 
$$
x(p,q)=  \sum_{n,m} x_{nm} e^{2\pi i (np+mq)}
$$ 
The two coordinates $(p,q)$ then become identified with the membrane surface.
After implementing this transformation within the Hamiltonian of the
large $N$ SQM, one recovers the light-cone gauge membrane world-volume
Hamiltonian \cite{dewit,townsend,banks}.
In the matrix theory philosophy, this membrane configuration only
needs to use a part of the total $U(N)$ gauge group. In particular,
if we consider a sector of the theory with a membrane configuration 
in the 7-8 direction, we can decompose the total $U(N)$ matrix as 
\ba
X^7 \is p + x^7(p,q) + \tilde{X}^7\nonu
X^8 \is q + x^8(p,q) + \tilde{X}^8\nonu
X^i  \is x^i(p,q) + \tilde{X}^i\nonumber
\ea
Here the first two terms describe the classical membrane and its
transverse fluctuations, while $\tilde{X}^i$ denotes the remaining 
part of $X^i$ that commutes with the fluctuating membrane background.

We can now copy the same procedure in the matrix string theory. In this
case the fields $X$ becomes $\sigma$ dependent, and thus at first sight
the membrane coordinate fields $x(p,q)$ also acquire this additional
dependence. However, just as for the D-particles, the presence of the
membrane configuration breaks the total $U(N)$ gauge symmetry to a subgroup.
In particular, this means the part of the eigenvalues that are occupied
by the membrane can no longer be permuted. The $p,q$ dependence does not
allow such an action. So, in a similar fashion as in the case of the D-particle
these eigenvalues necessarily correspond to short strings that can only
by in their ground states. Therefore the functions $x^i(p,q)$
do not acquire a $\sigma$-dependency. 

One should be able to analyze the interactions of the type IIA strings
with the D-membrane quite explicitly along the lines outlined in the
discussion of the D-particles. Again we expect that the ``long''
strings can now attach to ``short'' strings that constitute the
membrane, analogous to the way depicted in {\it fig.} 5. However, in this
case both end of the open string should be able to travel independently on
the D-membrane world-volume. We leave a more precise analysis of these
interactions for further study.

\bigskip
\bigskip

{\noindent \sc Acknowledgements}

We would like to thank S. Bais, T. Banks, N. Berkovits,
S. Das, F. Hacquebord, L. Motl, J-S. Park, W. Taylor, and B. de Wit for
useful discussions.  This research is partly supported by a Pionier
Fellowship of NWO, a Fellowship of the Royal Dutch Academy of Sciences
(K.N.A.W.), the Packard Foundation and the A.P. Sloan Foundation.

\bigskip
\bigskip

\newsubsection{Appendix}

\renewcommand{\theequation}{A.\arabic{equation}}
\setcounter{equation}{0}

In this appendix we derive the matrix string Hamiltonian starting from the
matrix theory formalism of \cite{banks}. In particular we wish to identify
the precise dependence on the string coupling constant. For a similar
derivation see \cite{motl}.

Starting point is the matrix theory Hamiltonian, written in eleven-dimensional
Planckian units with $\ell_p=1$ (we ignore numerical prefactors)
\be
{\cal H}
 = R_{11}\, \tr\Bigl(\,\Pi_i^2 + [X^i,X^j]^2 + \theta^T\gamma_i[X^i,\theta]
\,\Bigr).
\ee
We now compactify the 9th dimension on a circle of radious $R_9$. After the usual
T-duality we can identify $X^9$ with the covariant derivative $R_9 D_\sigma$,
where the coordinate  $\sigma$ runs from $0$ to $2\pi$. The conjugate momentum
will be identified with the electric field $E$ via $E=R_9\Pi_9$.
This gives the Hamiltonian (where $i=1,\ldots,8$ now labels the transverse coordinates)
\ba
{\cal H} \is  
{R_{11}\over 2\pi} \int  \! {d\sigma\over R_9} \, \tr\Bigl(\,\Pi_i^2 +  R_9^2\, (DX_i)^2
+ R_9 \,\theta^T D\theta \nonu && \qquad\qquad\ \  
+ \, {1\over R_9^2} \,E^2 + [X^i,X^j]^2 + \theta^T\gamma_i[X^i,\theta]
\,\Bigr).
\ea
One can rescale the coordinates as $X^i \ra R_9^{-1/2}X^i$ to find
\ba
{\cal H} \is  {R_{11} \over 2\pi}
 \int \! d\sigma \, \tr\Bigl(\,\Pi_i^2 +  (DX^i)^2 + \theta^TD\theta \nonu && 
\qquad \qquad\ \ 
+ \, {1\over R_9^3} \,(E^2 + [X^i,X^j]^2)
 + \,{1 \over R_9^{3/2}}\, \theta^T\gamma_i[X^i,\theta]
\,\Bigr).
\ea
Conventionally, M-theory is related to type IIA string theory via the
compactification of the 11th direction, which relates the string
coupling constant $g_s$ to $R_{11}^{3/2}$. To arrive at the matrix string
point of view, however, we now interchange the role of the 9th and the
11th direction by defining the string scale $\ell_s=\sqrt{\a'}$ and
string coupling constant $g_s$ in terms of $R_9$ and the 
11-dimensional Planck length
$\ell_p$
\be
R_9=g_s \ell_s,\qquad \ell_p =g_s^{1/3}\ell_s,
\ee
or equivalently $g_s = (R_9/\ell_p)^{3/2}$. From this we obtain the final  result
in string units $\ell_s = 1$
\ba
{\cal H} \is  {R_{11} \over 2\pi}
\int\! d\sigma \, \tr\Bigl(\,\Pi_i^2 + (DX^i)^2 + \theta^T  D\theta \nonu && 
\qquad\qquad \ \ 
+ \,{1\over g_s^2} \,(E^2 + [X^i,X^j]^2)
 + \,{1\over g_s} \,\theta^T\gamma_i[X^i,\theta]
\,\Bigr).
\ea
In this convention, $R_{11}$ normalizes the light-cone momentum $p_+$ via
\be
p_+ = N/R_{11}
\ee
and becomes infinite in the large $N$ limit. The normalization chosen
in the main text corresponds to total light-cone momentum $p_+=1$, so
that $R_{11}=N$ and the mass-shell relation reads $p_-=H$. In addition
we have absorbed this factor of $N$ into the definition of the
world-sheet time coordinate, so that the above Hamiltonian ${\cal H}$
is related to the world-sheet time generator $H=L_0 + \Lbar_0$ of the
CFT as ${\cal H}=N H$.

\renewcommand{\Large}{\large}

\end{document}